\documentclass{PoS}
\pdfoutput=1 

\title{The DMTPC  detector}

\ShortTitle{The DMTPC detector}

\author{\speaker{G.~Sciolla}\\
        Massachusetts Institute of Technology \\
        E-mail: \email{sciolla@mit.edu}\\ }

\author{ J.~Battat, T.~Caldwell, B.~Cornell, D.~Dujmic, P.~Fisher, S.~Henderson, R.~Lanza,  A.~Lee, J.~Lopez, A.~Kaboth, 
G.~Kohse,  J.~Monroe, T.~Sahin, R.~Vanderspek, R.~Yamamoto,  H.~Yegoryan \\
        Massachusetts Institute of Technology \\ }

\author{S.~Ahlen, D.~Avery, K.~Otis, A.~Roccaro, H.~Tomita \\
        Boston University\\ }

\author{A.~Dushkin, H.~Wellenstein \\ 
        Brandeis University\\ }

\abstract{
Directional Dark Matter detectors  have the potential of yielding an unambiguous observation of 
WIMPs even in presence of insidious background. 
In addition, by measuring the direction of the Dark Matter particles such detectors can 
discriminate between the various 
models that describe Dark Matter in our galaxy.  
The DMTPC detector is a novel directional DM detector consisting of  a  low-pressure CF$_4$ 
time projection chamber with optical readout. 
Recent measurements proved that this technology is able to reconstruct the energy, direction, 
and sense  of the low-energy  nuclear recoils 
produced by neutrons from a $^{252}$Cf  source, as well as efficiently reject  electromagnetic backgrounds. 
A 10-liter DMTPC  detector is ready for underground operation. A  1 m$^3$  detector, 
now in the design phase, will soon allow us to improve the existing limits of 
SD-interactions of WIMPs on protons by over one order of magnitude. 
}

\FullConference{Identification of dark matter 2008\\
		 August 18-22, 2008\\
		 Stockholm, Sweden}

\begin{document}

\section{Directional Dark Matter detectors }

Directional Dark Matter (DM) detectors~\cite{sciolla} are designed to detect 
the daily modulation in the direction of the Dark Matter wind~\cite{spergel}. 
The direction of the incoming WIMP is encoded in the direction of the 
nuclear recoil originated by the elastic scattering of the WIMPs with 
the atoms in the detector. A precision of 20--30 degrees can be obtained 
using a very dilute gas (40-100 torr) as a target material. 
Directional detectors offer unique ways to suppress a variety of backgrounds. 
The correlation between the energy and length of the nuclear recoil is 
very powerful in suppressing backgrounds from photons, electrons, and alpha particles. 
The determination of the direction of the nuclear recoils 
is used to discriminate between signal (correlated with the direction of the DM wind), 
neutron backgrounds (randomly distributed around the detector), and 
solar neutrino backgrounds~\cite{monroe}  (pointing back to the Sun). 

Finally, the determination of the direction of arrival of Dark Matter particles can 
discriminate between various DM halo distributions in our Galaxy\cite{alenazi},  making DM directional 
detectors unique observatories for underground WIMP astronomy.

\section{The DMTPC detector }  

The DMTPC detector is a low-pressure time projection chamber (TPC) with optical readout. 
The TPC is filled with tetrafluoromethane (CF$_4$) at a pressure of about 50 torr. 
At such a pressure, a typical collision of a WIMP with a gas molecule would cause a nucleus to recoil 
by about 2 mm. The ionization electrons produced by the recoiling nucleus drift in the gas along 
the electric field toward the amplification region. In the amplification process, 
scintillation photons are produced together with electrons. 
A CCD camera mounted above the cathode mesh records these photons to produce an image of the  
 nuclear recoil track as projected on the amplification plane. 
Because the energy loss is not uniform along the trajectory, one can determine the direction of the 
incoming WIMP (``head-tail'' measurement). 
An array of photomultipliers (PMTs) mounted above the cathode mesh 
measures the length of the recoil in the drift direction.

 CF$_4$ was chosen as  target material primarily because of its high content in fluorine. $^{19}$F is an ideal 
element~\cite{ellis} to detect spin-dependent interactions on protons, 
due to its large spin factor and isotopic abundance. 
In addition, CF$_4$ is an excellent detector material~\cite{Pansky,Christophorou}, with 
good scintillation properties~\cite{asher} and low transverse diffusion. 
Finally, CF$_4$ is non-flammable and non-toxic, and, therefore, safe to operate underground. 

Electron amplification is obtained by 
applying  a large potential difference ($\Delta$V = 0.6--1.1 kV) 
between two conductive woven meshes (or between a mesh and a  copper plate) separated by about 500 $\mu$m.
The meshes are made of  28 $\mu$m stainless steel or Cu wire,  with a pitch of 256 $\mu$m,  
which  determines the intrinsic spatial resolution of the detector. 
The CCDs provide 2-D readout at a very low  cost per channel, which 
makes directional detectors economically viable.

\section{Measurement strategy}   

The DMTPC detector  simultaneously measures   
the number of photons collected the CCD camera, the projection  of the recoiling nucleus along the amplification plane, 
the energy loss along the recoil track, the width and integral of the PMT signal,
the electronic signal produced on the amplification plane.   

The energy of the nuclear recoil can be independently reconstructed from 
the number of photons observed in the CCDs, the integral of the electronic signal 
produced on the amplification mesh, and the integral of the PMT signal. 
The track length of the recoiling nucleus is reconstructed by combining the measurement 
of the projection along the amplification plane (from pattern recognition in the CCD) 
and the projection along the direction of drift, determined from the width of the signal 
recorded in the PMTs. The sense of the recoil track can be determined by the Bragg curve, 
the characteristic variation of the energy deposit along the length of the track. 

The combination of these measurements reconstructs 
the energy, direction, and sense of  nuclear recoils from WIMPs, 
allowing for an  excellent rejection of the electromagnetic backgrounds.  
The gamma ray rejection factor, measured using a $^{137}$Cs source, is better than 2 parts per million.

The CCD images has a long (1 second) exposure. When a trigger is generated by the PMT or the electronic readout 
of the amplification plane, the CCD is read-out and the event is saved to disk. 
Otherwise, the CCD is simply reset, to minimize dead time. 


\section{ Current prototype and  R\&D results} 

\begin{figure*}[t]
\centering
\includegraphics[width=\textwidth]{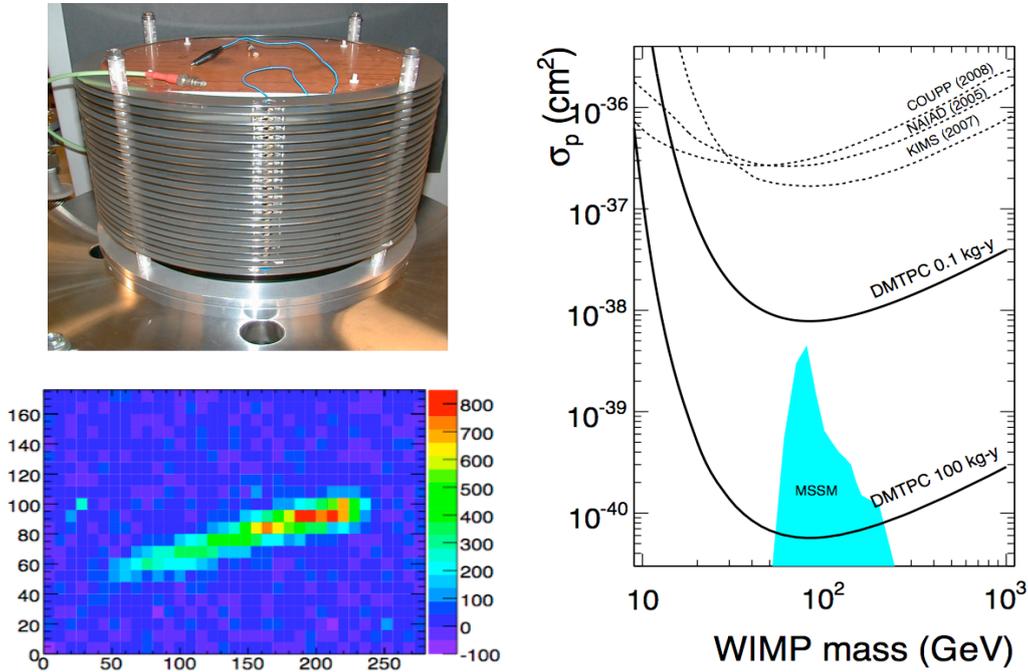}
\caption[]{
Top left: image of the 10-liter DMTPC prototype. 
Bottom left: image of a nuclear recoil from a low-energy neutron 
in  CF$_4$ at 75~torr. 
Right: expected sensitivity (90\% C.L.) to spin-dependent WIMP scattering
on protons for a DMTPC detector (solid lines) 
compared with published limits from other experiments (dashed lines). 
The shaded area shows the MSSM favored region.    
} \label{Fig1}
\end{figure*}

The current DMTPC prototype (Figure~\ref{Fig1}, top left) 
consists of two optically independent regions contained in one stainless steel vessel. Each region is a cylinder with 25 cm diameter and 25 cm height contained inside a field cage. 
The amplification is obtained by using a mesh-plate design. The detector is read-out by two CCD cameras, 
each imaging one drift region. The optical system uses two Nikon photographic lenses 
with f-number of 1.2 and focal length of 55 mm, 
and two Apogee U6 cameras equipped with  Kodak 1001E CCD chips. 
Because the total area imaged is $16\times16$~cm$^2$, the detector has an active volume of about 10 liters.

An $^{241}$Am source  producing 5.5 MeV alpha particles is used to 
study the gain of the detector as a function of the voltage 
in the amplification region and gas pressure. Typical gas gains are $\approx$ 10$^5$. 
We also  measure the transverse diffusion  as a function of the primary electrons drift distance 
These studies
show that the transverse diffusion is less then 1 mm for a drift distance of 25 cm~\cite{headtailpaper}.

The ability of the DMTPC detector to  determine the sense and direction of nuclear recoils has been evaluated by studying the recoil of fluorine nuclei in interaction with low-energy neutrons. 
For nuclear recoils with energy below 1 MeV, the energy deposition  decreases along the path of the recoil, 
allowing for the identification of  the ``head'' (``tail'') of the event by a smaller (larger) energy deposition. 
The initial measurements  
used 14 MeV neutrons from a deuteron-triton generator. 
The reconstructed recoils with energy between 200 and 800 keV allowed us to observe the 
 ``head-tail'' effect with a significance of  8 $\sigma$~\cite{headtailpaper}.

In subsequent measurements~\cite{meshpaper} we used lower energy neutrons generated by a 
$^{252}$Cf  source. To achieve a  2-D reconstruction of the nuclear recoils,  
a  mesh-based detector was utilized. 
Better sensitivity to lower energy thresholds was achieved by lowering the CF$_4$ pressure to 75 torr. 
 Figure~\ref{Fig1} (bottom left) shows a Cf-induced nuclear recoil reconstructed in the DMTPC detector.  
The neutron was traveling right to left.  The decreasing $dE/dx$ along the track direction, clearly visible in the image,   
proves that the detector is sensitive to the sense of the direction on an event-by-event basis. 
Measurements of the length and asymmetry of the recoil tracks as a function of their  energy   
show good agreement with the predictions of the simulation and 
prove good head-tail discrimination for recoils above 100 keV~\cite{meshpaper}. 
The ``head-tail'' discrimination is expected to extend to recoils above 50 keV 
when the  detector is operated at a pressure of 50 torr.

\section{Future detectors and expected sensitivity}

A 1-m$^3$ DMTPC detector is being designed. 
The apparatus consists of a stainless steel vessel of 1.3 m diameter and 1.2 m height.
Nine CCD cameras and nine PMTs are mounted on each of the top and bottom plates of the vessel, separated from the active 
volume of the detector by an acrylic window.  The detector consists of two optically separated regions. 
Each of these regions is equipped with a triple-mesh amplification device, mounted in between two symmetric drift regions.  
Each drift region has a diameter of 1.2 m and a height of 25 cm, for a total active volume of 1 m$^3$. 
A field cage made of stainless steel rings keeps the uniformity of the electric field within 1\% in the fiducial volume.
A gas system recirculates and purifies the CF$_4$. 


When operating the detector at a pressure of 50 torr at 21 degrees C, this module will contain 250 g of CF$_4$. 
Assuming a data-taking efficiency of 50\%, a one-year run will yield 45 kg-days.

The sensitivity of the DMTPC  detector to SD interactions of WIMPs on protons has been studied
for exposures of 0.1 and 100 kg-years (Figure~\ref{Fig1}, right). We assumed 
that the detector will be operated at a depth of 2,000 m.w.e. 
inside a 40 cm thick polyethylene neutron shielding, 
and the absence of internal backgrounds above the 50 keV threshold. 
This study shows that improvements of a factor 50 over the  existing measurements can be obtained by operating 
a 1-m$^3$ DMTPC detector for less than a year. 
A larger detector, with an active mass of $10^2$--$10^3$ kg, will be able to explore 
a significant portion of the MSSM parameter space. 
This detector is an ideal candidate for the DUSEL laboratory in South Dakota.

\section{ Conclusion } 
The measurement of the direction of the incoming WIMPs 
 by a directional Dark Matter detector can provide 
an unambiguous positive observation of Dark Matter particles 
as well as a unique tool for underground WIMP astronomy.  

Our DMTPC detector is designed to  measure the energy, direction, and sense of the nuclear recoils produced in elastic collisions 
of WIMPs in low-pressure CF$_4$ gas.  The combination of these measurements allows for excellent background 
rejection, while the use of an optical readout substantially reduces the costs,  making a ton-size directional 
detector economically viable. The choice of  CF$_4$  as the target material makes this detector well suited for 
studies of  spin-dependent interactions.

\section*{Acknowledgments}
The DMTPC project is supported by the DoE ADR program, 
the National Science Foundation, 
the Pappalardo Fellowship program, and the MIT Kavli Institute and  Physics Department.


\begin{thebibliography}{99}
\bibitem{sciolla} G.~Sciolla,  submitted to Modern Physics Letters A, arXiv:0811.2764 [astro-ph].
\bibitem{spergel} D.~N.~Spergel, Phys.\ Rev.\  D {\bf 37}, 1353 (1988).
\bibitem{monroe} J. Monroe and P. Fisher, Phys. Rev. D {\bf 76}, 033007 (2007).
\bibitem{alenazi}  M.~S.~Alenazi and P.~Gondolo,  Phys.\ Rev.\  D {\bf 77}, 043532 (2008).
\bibitem{ellis} R.~J.~Ellis and R.~A.~Flores, Phys. Lett. B 263, 259 (1991).
\bibitem{Pansky} A.~Pansky {\it et al.},  Nucl.\ Instrum.\ Meth.\  A {\bf 354}, 262 (1995).
\bibitem{Christophorou} L.~G.~Christophorou,  {\it et al.}, J. Phys. Chem. Ref. Data 25, 1341 (1996)  
\bibitem{asher} A.~Kaboth {\it et al.},  [DMTPC Collaboration], Nucl.\ Instrum.\ Meth.\  A {\bf 592}, 63 (2008).
\bibitem{headtailpaper} D.~Dujmic {\it et al.}, [DMTPC Collaboration],  Nucl.\ Instrum.\ Meth.\  A {\bf 584}, 327 (2008).
\bibitem{meshpaper} D.~Dujmic {\it et al.}, [DMTPC Collaboration], Astropart.\ Phys.\  {\bf 30}, 58 (2008).
\end{thebibliography}
\end{document}